%% file: main.tex
\documentclass{article}
\usepackage{spconf,amsmath,graphicx}
\usepackage{hyperref}
\usepackage{multirow}
\usepackage{stfloats}
\usepackage{booktabs}
\usepackage{bm}
\usepackage{enumitem}
\usepackage{threeparttable}

\title{DASA: Difficulty-Aware Semantic Augmentation for speaker verification}

\name{
\begin{tabular}{c}
Yuanyuan Wang$^{1,\ast}$, Yang Zhang$^{1}$, Zhiyong Wu$^{1,3,\dagger}$, Zhihan Yang$^{1}$, Tao Wei$^2$, Kun Zou$^2$, Helen Meng$^3$
\end{tabular}
\thanks{$^{\ast}$Work performed during an internship at Ping An Technology.}
\thanks{$^\dagger$Corresponding author.}
}
\address{
  $^1$ Shenzhen International Graduate School, Tsinghua University, Shenzhen, China\\
  $^2$ Ping An Technology, Shenzhen, China\\
  $^3$ The Chinese University of Hong Kong, Hong Kong SAR, China\\
  \small{\{wangyuan21, zhangy20\}@mails.tsinghua.edu.cn,  zywu@sz.tsinghua.edu.cn}
}

\begin{document}
\ninept
\maketitle
\begin{abstract}
%
%
Data augmentation is vital to the generalization ability and robustness of deep neural networks (DNNs) models. 
Existing augmentation methods for speaker verification manipulate the raw signal, which are time-consuming and the augmented samples lack  diversity. 
In this paper, we present a novel difficulty-aware semantic augmentation (DASA) approach for speaker verification, which can generate diversified training samples in speaker embedding space with negligible extra computing cost.
%
Firstly, we augment training samples by perturbing speaker embeddings along semantic directions, which are obtained from speaker-wise covariance matrices. 
Secondly, accurate covariance matrices are estimated from robust speaker embeddings during training, so we introduce difficulty-aware additive margin softmax (DAAM-Softmax) to obtain optimal speaker embeddings. 
Finally, we assume the number of augmented samples goes to infinity and derive a closed-form upper bound of the expected loss with DASA, which achieves compatibility and efficiency.
Extensive experiments demonstrate the proposed approach can achieve a remarkable performance improvement.
The best result achieves a 14.6\% relative reduction in EER metric on CN-Celeb evaluation set.


\end{abstract}

\input{01_introduction}
\input{02_method}

\input{03_experiments}

\input{04_conclusions}

\vfill\pagebreak

\bibliographystyle{IEEEbib}
\bibliography{strings,refs}

\end{document}

%% file: 01_introduction.tex
\section{introduction}
\label{sec:introduction}
Speaker verification is to compare two speech utterances and verify whether they are spoken by the same speaker \cite{7298570}.
In recent years, deep neural networks (DNNs) have significantly boosted advancements in speaker verification \cite{6854363,he2016deep,snyder2018x,desplanques2020ecapa,zhang2022mfa}.  
Now speaker verification has become an important technology in our daily life, such as biometric identification or smartphone control. However, current speaker verification systems are still unsatisfactory in real industrial scenarios, 
such as video analysis with non-standard ambient sound, channel number, and speech emotion.
Data insufficiency is one of the most critical challenges to the robustness of performance in complex scenarios.

To address the problem of data insufficiency, data augmentation is an essential technique to increase the diversity and quantity of training samples \cite{yamamoto2019speaker}. 
Conventional data augmentation techniques for speaker verification, e.g., additive noises, reverberation, and speed perturbation \cite{snyder2018x,ko2015audio}, are raw signal-level data augmentation. 
Wang \textit{et al.}~\cite{wang2020investigation} investigate SpecAugment~\cite{park19e_interspeech} and augment data by masking the spectrogram during training. 
The diversity of augmented samples generated by these methods is inherently limited by the direct manipulation of the raw audio~\cite{ko2015audio}. 
Besides, these methods will also introduce huge computing cost and I/O time for augmentation. 

Recently, 
Deep generative models~\cite{yang2018generative,wu2019data}, e.g., Generative Adversarial Networks (GANs), Variational Autoencoder (VAE), have been introduced to learn the distribution of noisy speaker embeddings and generate new embeddings from the generative models learned distribution. 
However, these methods utilize complex deep generative models to explicitly augment samples, which significantly slow down the training process of recognition models. 
Xun \textit{et al.}\cite{9362090} sample noise from the pure noise distribution and directly add it to clean embeddings to generate augmented embeddings. But the distribution is derived and specialized on extra noise datasets.


These problems can potentially be solved by the implicit semantic data augmentation (ISDA) approach \cite{wang2019implicit,9332260}.
Noise data, auxiliary model, and model modification are no longer required, and therefore, it eliminates extra computation and time cost.
ISDA augments training data by translating speaker embeddings towards meaningful semantic directions, which are sampled from a zero-mean normal distribution with the dynamically estimated covariance \cite{li2021metasaug}. 
Importantly, ISDA estimates speaker-wise covariance matrices according to speaker embeddings. We find that ISDA performs sub-optimally in speaker verification tasks, because it is based on softmax training, unable to acquire the optimal embedding to estimate the appropriate covariance matrices. 

In this paper, we propose a novel difficulty-aware semantic augmentation (DASA) approach to augment the training data at deep speaker embedding level rather than the raw signal level, from the point that more accurate covariance matrices require optimal embeddings. Without changing the model structure, the loss function plays a crucial role in speaker embeddings. 
We first use additive margin softmax (AM-Softmax) \cite{8331118} loss to reduce the intra-class variation and increase the inter-class difference \cite{8331118,9023039}. 
Furthermore, AM-Softmax assumes that all the speakers have the same inter-class difference. 
Difficult samples are highly similar to other speakers and have insufficient discrimination, so their inter-class difference should be larger.
Therefore, we are inspired by \cite{https://doi.org/10.48550/arxiv.2104.06094} and introduce difficulty-aware AM-Softmax (DAAM-Softmax) which sets inter-class difference according to sample difficulty.  
DAAM-Softmax can solve the over-optimization of easy samples and under-optimization of difficult samples, so it gets the optimal embeddings to estimate covariance matrices of DASA. 
Finally, we assume infinite sampling directions and derive an upper bound of expected loss with DASA, which can be simply adopted by most models with little computation and time cost.
In summary, our contributions are as follows:

\begin{itemize}[itemsep=0pt,topsep=0pt,parsep=0pt,leftmargin=10pt]
\item \textbf{Low computing cost}: Instead of augmenting the speech by processing the raw signal, DASA performs meaningful semantic perturbation in speaker embedding space with little extra computing cost.
\item \textbf{Good compatibility}: The proposed DASA approach can combine with traditional data augmentation and is compatible with most models to achieve better recognition performance.
\item \textbf{Outstanding improvement}: Extensive experiments 
conducted on VoxCeleb and CN-Celeb demonstrate the proposed method can obtain remarkable performance improvement.
\end{itemize}

%% file: 02_method.tex
\vspace{-1pt}
\section{Difficulty-Aware Semantic Augmentation}
\label{sec:method}
\begin{figure}[htb]
\centering
\includegraphics[width=8.5cm]{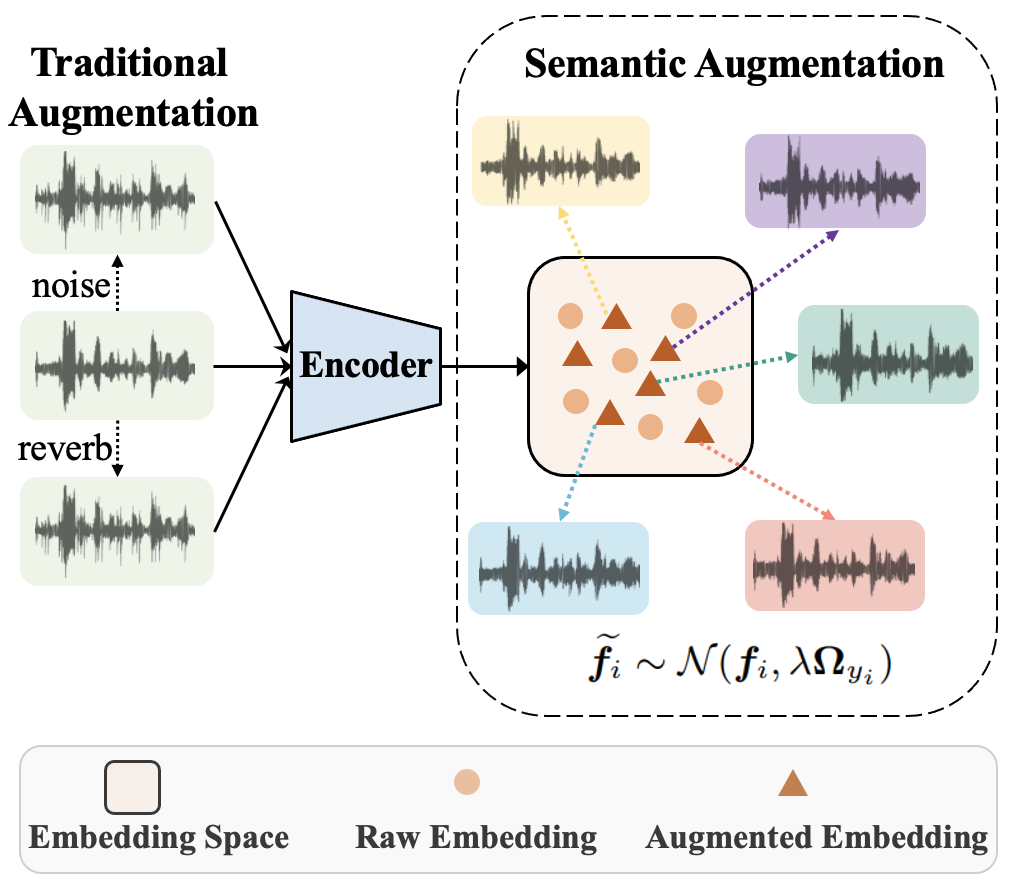}
\vspace{-5pt}
\caption{\textbf{The comparison between conventional data augmentation (left) and proposed difficulty-aware semantic augmentation (right).} 
Conventional data augmentations directly perform on the raw signal level, such as adding noise and reverberation, whereas the DASA augments the training data at the speaker embedding level.
DASA can exactly estimate speaker-wise covariance matrices in training with DAAM-Softmax, and these covariance matrices are used to establish a zero-mean normal distribution. Then DASA generates new instances by changing speaker embeddings to semantic transformation directions sampled from the normal distribution.  
Augmented diversified instances are represented by speech rectangles with different colors. After that, to improve efficiency, we assume an infinite number of sampling and derive a closed-form upper bound of the expected loss with DASA.}
\label{fig:structure}
\vspace{-4pt}
\end{figure}

In this section, we present the difficulty-aware semantic augmentation (DASA) for speaker verification. 
The comparison between conventional data augmentation and the proposed DASA is shown in Fig.\ref{fig:structure}. 
We assume the training set is $S=\{(\bm{u}_i,y_i)\}_{i=1}^N$, and $y_i \in \{1,\ldots,C\}$ is the label of the i-th utterance $\bm{u}_i$ over C speaker categories. The vector $\bm{f}_i=[f_{i1},\ldots,f_{iF}]^T$ indicates the F-dimensional deep embeddings of $\bm{u}_i$ learned by deep neural network D. 

\vspace{-10pt}
\subsection{Implicit Semantic Data Augmentation}
Here, we revisit the ISDA proposed by \cite{wang2019implicit}. 
We define that speaker-wise covariance matrix $\bm{\Omega}_{y_i}$ is computed from all embeddings of training samples in class $y_i$, so C speakers have C covariance matrices during training. We can then randomly sample from a zero-mean normal distribution $\mathcal N(0, \lambda \bm{\Omega}_{y_i})$ to obtain vectors representing semantic directions of $\bm{f}_i$. The augmented embedding $\widetilde{\bm{f}}_i$ is generated by Eq.(\ref{sample}), where $\lambda$ is a positive number that controls the strength of ISDA.
\begin{equation}\label{sample}
\begin{aligned}
\widetilde{\bm{f}}_i \sim \mathcal{N}(\bm{f}_i, \lambda\bm{\Omega}_{y_i})
\end{aligned}
\end{equation}

Firstly, we explicitly augment each embedding $\bm{f}_i$ for M times. Then through the softmax layer, the cross entropy (CE) loss function is calculated as Eq.(\ref{celoss}), where \textbf{w} and \textbf{b} are the weight matrices and biases of the last fully connected layer, respectively.
\begin{equation}
\label{celoss}
\begin{aligned}
L_M &= \frac{1}{N}\sum_{i=1}^N\frac{1}{M}\sum_{k=1}^M-log(\frac{e^{\bm{w}_{y_i}^T \bm{f}_i^k+b_{y_i}}}{\sum_{j=1}^C e^{\bm{w}_j^T \bm{f}_i^k+b_j}})
\end{aligned}
\end{equation}

When considering that $M \to\infty$, an easy-to-compute upper bound can be derived for the loss function. 
The expectation of the CE loss is calculated as all possible augmentation embeddings and uses Jensen's inequality $E(logX)\leq logE(X)$ to approximate:
\begin{equation}\label{expect}
\begin{aligned}
L_\infty 
&= \frac{1}{N}\sum_{i=1}^{N}E_{\widetilde{\bm{f}}_i}(log(\sum_{j=1}^Ce^{(\bm{w}_j^T-\bm{w}_{y_i}^T)\widetilde{\bm{f}}_i+(b_j-b_{y_i})})) \\
&\leq \frac{1}{N}\sum_{i=1}^{N}log(\sum_{j=1}^CE_{\widetilde{\bm{f}}_i}(e^{(\bm{w}_j^T-\bm{w}_{y_i}^T)\widetilde{\bm{f}}_i+(b_j-b_{y_i})}))
\end{aligned}
\end{equation}

According to Eq.(\ref{sample}), we can deduce Eq.(\ref{gaussian}), where $\Delta\bm{w}=(\bm{w}_j^T-\bm{w}_{y_i}^T)$, 
$\Delta{b}=(b_j-b_{y_i})$ and $\bm{\Phi}=(\bm{w}_j^T-\bm{w}_{y_i}^T) \bm{\Omega}_{y_i} (\bm{w}_j-\bm{w}_{y_i})$.
\begin{equation}\label{gaussian}
\begin{aligned}
\Delta\bm{w}\widetilde{\bm{f}}_i+\Delta{b} \sim \mathcal{N}(\Delta\bm{w}\bm{f}_i+ \Delta{b}, \lambda \bm{\Phi})
\end{aligned}
\end{equation}

Combined with the Moment Generating Function Eq.(\ref{MGF}), we can derive an easily computable upper bound as the final loss function Eq.(\ref{soft-isda}). Without explicitly augmented samples, ISDA can be performed more efficiently by calculating the upper bound of the loss. 
\begin{equation}\label{MGF}
\begin{aligned}
E[e^{tX}]=e^{t\mu+\frac{1}{2}\sigma^2 t^2}, X \sim N(\mu, \sigma^2)
\end{aligned}
\end{equation}
\begin{equation}\label{soft-isda}
\begin{aligned}
L_{\infty (ISDA)}  
&\leq
\frac{1}{N}\sum_{i=1}^{N}log( 
\sum_{j=1}^Ce^{\Delta\bm{w}\bm{f}_i+\Delta{b}+\frac{1}{2}\lambda \bm{\Phi}})  
\end{aligned}
\end{equation}

\vspace{-8pt}
\subsection{AM-Softmax with Semantic Augmentation}
The above formula Eq.\ref{soft-isda} performs poorly in speaker verification, since it is based on the softmax function. 
AM-Softmax can better separate different speakers and make speaker embeddings from the same speaker more compact by introducing the angular margin into the softmax. So it can get better embeddings to accurately estimate covariance matrices $\bm{\Omega}_{y_i}$.

In this section, we will derive the upper bound of expected AM-Softmax loss with semantic augmentation. The formula for AM-Softmax loss commonly used in speaker verification \cite{8331118} can be formulated as follows, where both $\bm{w}$ and $\bm{f}$ are normalized.
\vspace{-1pt}
\begin{equation}\label{AM-Softmax}
\begin{aligned}
L_{AMS}
&=-\frac{1}{N}\sum_{i=1}^Nlog\frac{e^{s\cdot(cos\theta_{y_i}-m)}}{e^{s\cdot(cos\theta_{y_i}-m)}+\sum_{j=1, j\neq y_i}^Ce^{s \cdot cos\theta_j}}\\
&=-\frac{1}{N}\sum_{i=1}^Nlog\frac{e^{s(\bm{w}_{y_i}^T\bm{f}_i-m)}}{e^{s(\bm{w}_{y_i}^T\bm{f}_i-m)}+\sum_{j=1, j\neq y_i}^Ce^{s \bm{w}_j^T\bm{f}_i}}
\end{aligned}
\end{equation}

Similar to formulas Eq.(\ref{celoss}) and Eq.(\ref{expect}), we also sample $M \to\infty$ times and approximate the upper bound of the expected loss:
\vspace{-1pt}
\begin{equation}\label{daisda_expect}
\begin{aligned}
L_{A} 
&=\frac{1}{N}\sum_{i=1}^{N}E_{\widetilde{\bm{f}}_i}(log(1+\sum_{j=1,j\neq y_i}^Ce^{s \Delta\bm{w}\widetilde{\bm{f}}_i+sm})) \\
&\leq \frac{1}{N}\sum_{i=1}^{N}log(1+\sum_{j=1,j\neq y_i}^CE_{\widetilde{\bm{f}}_i}(e^{s \Delta\bm{w}\widetilde{\bm{f}}_i+sm}))
\end{aligned}
\end{equation}

According to equation Eq.(\ref{sample}), we can obtain the following Gaussian distribution:
\begin{equation}\label{isda-gaussian}
\begin{aligned}
s\Delta\bm{w}\widetilde{\bm{f}}_i+ sm\sim \mathcal{N}( s\Delta\bm{w}\bm{f}_i+sm, 
\lambda \bm{\Phi}s^2)
\end{aligned}
\end{equation}

From formula Eq.(\ref{MGF}), the upper bound of the expected AM-Softmax loss with semantic augmentation can be obtained:
\begin{equation}\label{amsoft_isda}
\begin{aligned}
L_{A}
&\leq \frac{1}{N}\sum_{i=1}^{N}log(1+ 
\sum_{j=1,j\neq y_i}^Ce^{s \Delta\bm{w} \bm{f}_i+sm + \frac{1}{2} \lambda \bm{\Phi} s^2)})
\end{aligned}
\end{equation}

\subsection{Difficulty-aware Semantic Augmentation}
All the speakers have the same inter-class difference in AM-Softmax, so the margin in the AM-Softmax loss is set to be a fixed value. 
But in fact, the angle between the speaker embedding and the center of the speaker category is different for each sample \cite{zhou2020dynamic}. 
When the angle is larger, it indicates that the samples are more difficult to verify. A larger margin should be set to increase inter-class difference. 
For simple training utterances, setting a slightly smaller margin can distinguish well. 
So we introduce difficulty-aware AM-Softmax (DAAM-Softmax) as shown in Eq.(\ref{DA-AM-Softmax}).
Harder samples get larger margins, thus achieving the purpose of difficulty-aware discriminative learning. 
\vspace{-3pt}
\begin{equation}\label{da}
\begin{aligned}
DA=\frac{1-cos\theta_{y_i}}{2}
\end{aligned}
\end{equation}
\begin{equation}\label{DA-AM-Softmax}
\begin{aligned}
L_{DAAM} 
&=-\frac{1}{N}\sum_{i=1}^Nlog\frac{e^{s(\bm{w}_{y_i}^T\bm{f}_i-m\cdot DA)}}{e^{s(\bm{w}_{y_i}^T\bm{f}_i-m\cdot DA)}+\sum_{j=1, j\neq y_i}^Ce^{s \bm{w}_j^T\bm{f}_i}}
\end{aligned}
\end{equation}

The $DA$ makes the coefficient in the range of $[0,1]$, and ensures difficulty is negatively correlated with logits. It will focus more on difficult utterances that usually lead to worse results \cite{https://doi.org/10.48550/arxiv.2104.06094}. 
This enables further accuracy improvements across all training samples which can generate optimal deep embeddings. Therefore, the covariance matrix $\bm{\Omega}_{y_i}$ estimated with optimal embeddings is more accurate. 
Based on our DAAM-Softmax, the upper bound of the expected loss with the proposed difficulty-aware semantic augmentation ($\bm{DASA}$) is Eq.(\ref{da_amsoft_isda}). 
Therefore, DASA can be easily applied to most deep models as a novel robust loss function.
\begin{equation}\label{da_amsoft_isda}
\begin{aligned}
L_{\infty (DASA)}
&\leq \frac{1}{N}\sum_{i=1}^{N}log(1+ 
\sum_{j=1,j\neq y_i}^Ce^{s \Delta\bm{w} \bm{f}_i+sm\cdot DA + \frac{1}{2} \lambda \bm{\Phi} s^2)})
              \end{aligned}
\end{equation}

%% file: 03_experiments.tex
\section{experimental setup}
\label{sec:exp_res}
\vspace{-8pt}
\subsection{Dataset}

VoxCeleb1\&2 \cite{Nagrani17,Chung18b} and CN-Celeb \cite{fan2020cn,li2022cn} 
are used in our experiments. 
VoxCeleb is an audio-visual dataset consisting of 2,000+ hours of short clips of human speech.
The development set of VoxCeleb2 containing 5994 speakers is used to train models. We use VoxCeleb1 test set to evaluate system performance.
CN-Celeb is a more challenging large-scale text-independent  speaker recognition dataset containing 2800 speakers for network training and 200 speakers for system evaluation. 


\vspace{-8pt}
\subsection{System Implementation}

\textbf{Network.} The experimental systems are based on  WeSpeaker~\footnote{https://github.com/wenet-e2e/wespeaker}.
To sufficiently verify the effectiveness of our method, we use ResNet34~\cite{he2016deep} and ECAPA-TDNN~\cite{desplanques2020ecapa} as speaker models, which accept 80-dimensional
Fbanks as input 
and extract 256-dimensional speaker embeddings. 

\noindent \textbf{Implementation details.}
The initial and final learning rates are 0.1 and 5e-5, which are decreased with an exponential schedule. 
We train all models for 150 epochs using an SGD optimization algorithm with a Nesterov momentum.
We set the weight decay as 1e-4 and the momentum as 0.9. The batch size is 128.
In most experiments, we perform traditional data augmentation online, including adding noise and reverberation but no speed perturbation. 
Each training utterance has a probability of 0.6 for conventional data augmentation~\cite{https://doi.org/10.48550/arxiv.2206.11699}.

In addition, we adopt a simple but effective deferred optimization schedule~\cite{NEURIPS2019_621461af}. In the first 60 epochs, we train models with DAAM-Softmax loss for learning good embeddings. 
In later epochs, we estimate the covariance matrix for DASA with
$\lambda=(t/T)\cdot \lambda_0$, where $t$ is the current iteration.


\vspace{-3pt}
\section{Results and Analyses}
\vspace{-3pt}
\subsection{Results on VoxCeleb}
\vspace{-6pt}
\begin{table}[htbp]
\vspace{-14pt}
\centering
\setlength{\tabcolsep}{1.8mm}
\caption{Results on VoxCeleb}
\vspace{1.5pt}
\begin{tabular}{cccc}
\toprule[1pt]
\multirow{2}{*}{\textbf{Loss}} & \multirow{2}{*}{\textbf{Hyperparams}} & \multicolumn{2}{c}{\textbf{VoxCeleb1-test}} \\
                               &                                           & \textbf{EER}(\%)      & \textbf{minDCF}    \\
\hline
Softmax                        & /                                         & 1.760              & 0.210               \\
ISDA \cite{wang2019implicit}                        & $\lambda_0=7$                                         & 1.191              & 0.141               \\
AM-Softmax \cite{8331118}                     & $s=32, m=0.2$                               & 0.936             & 0.093              \\
DAM-Softmax \cite{zhou2020dynamic}                    & $s=32, m=0.2$                               & 1.032             & 0.107              \\
\hline
\multirow{2}{*}{\textbf{DASA(Ours)}}              & $\lambda_0=4$                                  & 0.927             & 0.094              \\
              & $\lambda_0=3$                                  & \textbf{0.904}    & 0.098   \\
\bottomrule[1pt]
\end{tabular}
\begin{tablenotes}
\footnotesize
\item $^* \lambda_0$ indicates the strength of DASA. The scaling factor (s) is 32, and the margin (m) is 0.2 for all experiments except softmax and ISDA.
\end{tablenotes}
\label{res_on_vox}
\vspace{-10pt}
\end{table}
\begin{table*}[ht]
\centering
\setlength{\tabcolsep}{8.5mm}
\caption{Results on CN-Celeb}
\vspace{1.5pt}
\begin{tabular}{ccccc}
\toprule[1pt]
\multirow{2}{*}{\textbf{Model}} & \multirow{2}{*}{\textbf{Loss}}      & \multirow{2}{*}{\textbf{Hyperparams}} & \multicolumn{2}{c}{\textbf{CN-Celeb-Eval}} \\
                                 &                            &                                  & \textbf{EER}(\%)               & \textbf{minDCF}          \\
\hline
\multirow{7}{*}{ECAPA-TDNN}  & Softmax                    & /                                & 10.641                & 0.485        \\
                                 & AM-Softmax \cite{8331118}                 & $s=32, m=0.2$                      & 8.877                 & 0.435           \\
                            & DAM-Softmax \cite{zhou2020dynamic}                 & $s=32, m=0.2$                      & 8.348                 & 0.438          \\
    \cline{2-5}
                                 & \multirow{5}{*}{\textbf{DASA(Ours)}}          
                        & $\lambda_0=0.15$             & 8.278                 & 0.438   \\
                                  & 
                        & $\lambda_0=DA$               & 8.255                 & 0.443           \\
                     &           & $\lambda_0=0.2$               & 8.235                 & \textbf{0.432}           \\
                                 &  & $\lambda_0=0.1$              & \underline{8.161}        & 0.437           \\
                                 &  & $\lambda_0=DY$               & \textbf{8.021}        & 0.443           \\
\hline


\multirow{8}{*}{ResNet34}    & Softmax                    & /                                & 9.367                      &   0.429           \\
                                 & AM-Softmax \cite{8331118}                & $s=32, m=0.2$                      & 8.404                 & 0.416           \\
                    & DAM-Softmax \cite{zhou2020dynamic}                 & $s=32, m=0.2$                      & 7.977                 & 0.407           \\
        \cline{2-5}
                                 &
              \multirow{5}{*}{\textbf{DASA(Ours)}}          & $\lambda_0=DY$               & 7.784                 & \textbf{0.403}  
              \\
                                 &     & $\lambda_0=DA$               & 7.577                 & 0.406          \\
                                 &    & $\lambda_0=0.1$              & 7.568                 & 0.406           \\
                                 
                                 &   & $\lambda_0=0.2$             & 7.413       & 0.412        \\
                                 &   & $\lambda_0=0.15$             & \textbf{7.379}        & \underline{0.404}        \\
\bottomrule[1pt]
\end{tabular}
\label{res_on_cnc}
\vspace{-13pt}
\end{table*}
In the first experiment, we train ECAPA-TDNN model on VoxCeleb2, and test performance on VoxCeleb1 test set and results are shown in Table \ref{res_on_vox}.
ISDA performs worse than AM-Softmax on speaker verification, because it is based on softmax loss. So we will not conduct the ISDA experiment later. DAM-Softmax in the fourth line comes from the dynamic-margin softmax 
proposed in 
\cite{zhou2020dynamic}. 
It proposes to multiply the fixed margin by a dynamic variable as follows:
\begin{equation}
\label{dy}
\begin{aligned}
DY=\frac{e^{(1-cos\theta_{y_i})}}{\gamma}
\end{aligned}
\end{equation}
where $\gamma$ is set to 2. Since DY is similar to Eq.(\ref{da}), our DASA is 
compared with DAM-Softmax in all subsequent experiments.
The result of DAM here is slightly worse than AM-Softmax, which may be due to the large coefficient caused by the exponential form of DY.

The last two rows are the proposed DASA in Table \ref{res_on_vox}, where $\lambda_0$ represents the strength of DASA. DASA outperforms the AM-Softmax (baseline) and DAM-Softmax methods, and the performance is optimal when $\lambda_0$ is 3.

\vspace{-7pt}
\subsection{Results on CN-Celeb}
\vspace{-3pt}
Compared to VoxCeleb, CN-Celeb is a more complex and challenging dataset \cite{fan2020cn}. This is because speakers have 11 different genres of utterances, which leads to significant variation in speaking styles.
In addition, most short utterances recorded at different times with different devices involve real-world noise and perceptible voices in the background. 
To meet the demand for more real application scenarios, we train the ECAPA-TDNN and ResNet34 models on the CN-Celelb training set and test the performance on the CN-Celeb test set in this experiment. 

The hyperparameters of AM-Softmax have been carefully tuned to achieve outstanding performance and we experiment on this basis. 
Results in table \ref{res_on_cnc} show that DASA performs significantly better than AM-Softmax in both models, where $\lambda_0=DA$ and $\lambda_0=DY$ mean to change the strength of semantic augmentation with Eq. (\ref{da}), (\ref{dy}) dynamically. 
Bold and underline indicate the optimal and sub-optimal results, respectively.
DASA under both ECAPA-TDNN and ResNet34 models obtain 9.6\% and 12.2\% relative reduction in EER compared with AM-Softmax when evaluated on the CN-Celeb test set, respectively. In the case of ResNet34 and $\lambda_0=0.15$, both EER and minimum Detection Cost Function (minDCF) are almost optimal. 

Remarkable improvements in the experiments further demonstrate the effectiveness of DASA. 
Furthermore, compared with the experiments on VoxCeleb in Table \ref{res_on_vox}, 
the improvement on CN-Celeb is more apparent, which sufficiently verifies that DASA can pay more attention to difficult samples and performs better on complex trials of real industrial scenarios. 

\vspace{-7pt}
\subsection{Ablation Study}
\vspace{-1pt}
\label{sec:ablation_study}
In this section, we conduct ablation studies to explore the effect of each component in DASA. 
Table \ref{res_ablation} shows the results of ECAPA-TDNN on CN-Celeb, with similar trends on other models and datasets. 
Compared with the proposed DASA, we only remove SA and DA in the `w/o SA' and `w/o DA' lines, respectively. 
The `w/o SA' uses only DAAM-Softmax as shown in Eq.(\ref{DA-AM-Softmax}), and so does not include any hyperparameters for strength of DASA. 
The `w/o DA' uses the Eq.(\ref{amsoft_isda}) with $\lambda_0=0.1$, which allows for a fair comparison with our DASA. 
The results denote that both DA and SA can improve performance, with improvement of DA being slightly more apparent. 
The reduction of EER in the `w/o SA' row verifies that DA helps the model learn better embeddings.
The comparison of DASA and the `w/o DA' line shows that performance of SA is significantly improved with DA, indicating better embeddings are beneficial for SA to estimate covariance matrices.

\begin{table}[htbp]
\vspace{-8pt}
\centering
\setlength{\tabcolsep}{1.8mm}
\caption{Ablation study of ECAPA-TDNN on CN-Celeb}
\vspace{1pt}
\begin{tabular}{cccc}
\toprule[1pt]
\multirow{2}{*}{\textbf{Loss}} & \multirow{2}{*}{\textbf{Hyperparams}} & \multicolumn{2}{c}{\textbf{CN-Celeb-Eval}}   \\
                      &                                  &\textbf{EER}(\%)            &\textbf{minDCF}            \\
\hline
\multirow{2}{*}{\textbf{DASA(Ours)}}     & $\lambda_0=0.1$                & \underline{8.161}                  &0.437                 \\  

     & $\lambda_0=DY$                & \textbf{8.021}                  &0.443                 \\ 
\hline
w/o SA         & /
& 8.433              & 0.44           \\

w/o DA       & $\lambda_0=0.1$                       & 8.632              & 0.439             \\
\hline
w/o aug     & $\lambda_0=0.15$                &  9.175                  & 0.479                \\  
w/o aug, w/o DASA     & /                & 10.739                   & 0.485                \\ 

\bottomrule[1pt]
\end{tabular}
\begin{tablenotes}
\footnotesize
\item $^*$w/o aug means no traditional data augmentation.
\end{tablenotes}
\label{res_ablation}
\vspace{-16pt}
\end{table}

Since all the experiments in Table \ref{res_on_vox} and Table \ref{res_on_cnc} are performed with traditional data augmentation, we conduct experiments to eliminate the impact. 
Comparing the fifth row with the sixth row shows that the performance of DASA 
is relatively improved by 14.6\% without traditional augmentation, which exceeds the performance improvement with traditional augmentation. 
Furthermore, this indicates that DASA and traditional augmentation are complementary and can be combined to achieve higher performance.

More importantly, we calculate the average running time of an epoch as extra cost on Intel(R) Xeon(R) Silver 4210R CPU (2.40GHz) and GeForce RTX 3080. Compared with AM-Softmax, the cost introduced by traditional augmentation is about 213.3\% due to the limitation of I/O.
However, the additional cost of DASA on ECAPA-TDNN and ResNet34 is only 7.4\% and 8.5\%, respectively.


%% file: 04_conclusions.tex
\section{conclusions}
\vspace{-3pt}
\label{sec:conclu}
%
%
%
In this study, we present difficulty-aware semantic augmentation (DASA), a novel data augmentation approach for speaker verification. 
Different from conventional augmentation methods, which perform on raw speech signal level, the proposed DASA augments the training data on deep speaker embedding level without noticeable extra computing cost.
Besides, DASA could be an ideal complement to existing data
augmentation and be applied to various networks.
Extensive experiments on VoxCeleb and CN-Celeb show that DASA is effective and performs better on more realistic and challenging trials.
In the future, our method can hopefully be extended to other margin losses similar to AM-Softmax.

\textbf{Acknowledgement}: This work was supported by National Natural Science Foundation of China (62076144), CCF-Tencent Open Research Fund (RAGR20210122) and Shenzhen Science and Technology Program (WDZC20200818121348001).